# Compromising a Medical Mannequin

**William Bradley Glisson**
University of South Alabama
bglisson@southalabama.edu

**Todd Andel**
University of South Alabama
tandel@southalabama.edu

**Todd McDonald**
University of South Alabama
jtmcdonald@southalabama.edu

**Mike Jacobs**
University of South Alabama
mjacobs@southalabama.edu

**Matt Campbell**
University of South Alabama
mattcampbell@southalabama.edu

**Johnny Mayr**
University of South Alabama
jm1003@jagmail.southalabama.edu

## Abstract

Medical training devices are becoming increasingly dependent on technology, creating opportunities that are inherently conducive to security breaches. Previous medical device research has focused on individual device security breaches and the technical aspects involved with these breaches. This research examines the viability of breaching a production-deployed medical training mannequin. The results of the proof of concept research indicate that it is possible to breach a medical training mannequin in a live environment. The research contribution is an initial empirical analysis of the viability of compromising a medical training mannequin along with providing the foundation for future research.

**Keywords**

Medical, Security, Health Care Equipment, Training

## Introduction

The increasingly aggressive pervasive integration of technology into medical arenas creates new frontiers for malicious attacks with potentially serious, long-run cyber physical consequences. A recent Federal Bureau of Investigation (FBI) Cyber Division public notification states that "the health care industry is not technically prepared to combat against cyber criminals' basic cyber intrusion tactics" (Federal Bureau of Investigation 2014). They also state that intrusion activities are likely to increase against healthcare systems due to financial motives. In concert with this information a SANS report indicates that critical information assets in the healthcare industry are "poorly protected and often compromised" (Filkins 2014).

According to the U.S. Food and Drug Administration (FDA), there are currently no specific patient injuries or deaths associated with attacks on medical devices (2013b). However, the FDA has recently issued a warning stating that such devices can be vulnerable to cyber-based attacks (Wilson 2013). The GAO report (United States Government Accountability Office Report 2012) affirms that information security risks for medical devices are a new field for both federal regulators and the academic research community. In fact, hospitals and researchers have reported a growing number of exploitable vulnerabilities and possible intentional failures in the last five years (Basu 2013; Fu and Blum 2013; Industrial Control Systems Cyber Emergency Response Team 2013; Lee 2014; Lemos 2013; Pauli 2013; Weaver 2013)

The Government Accountability Office report (2012) concludes that the FDA faces challenges in how to identify medical device failures from intentional or malicious activities and that current policy and procedures are inadequate given the real security threats that exist. Fu and Blum (2013) also highlight the issue that federal reporting under FDA's Manufacturer and User Facility Device Experience (MAUDE) database does not adequately capture security-based failures or malfunctions for medical devices. In response, the FDA has released new guidance (2013a) (2013b) to address these concerns and has even encouraged researchers to disclose medical device vulnerabilities (Storm 2013). Neely (2013) points out that the guidance is too high-level to be useful and doesn't leave hospitals and device manufactures with actionable steps to instigate improvements.

It is reasonable to hypothesize that if the operational theater and/or a variety of medical devices typically found in a hospital are potentially vulnerable to cyber-attacks, then the training facilities could also, theoretically, be vulnerable to cyber-attack. The impact of a successful attack on a training facility, potentially, has much larger and longer lasting ripple effects in that subtle modifications could go undetected and yet influence training classes of medical professionals to incorrectly assess situations based on inaccurate feedback from medical devices.

This line of thought lead to the hypothesis that mannequins used in medical simulation training environments are at risk of cyber-attacks. The hypothesis raised several subsidiary research questions that were explored in order to address the hypothesis that included: Which components of the medical mannequin are vulnerable to attack? Can successful attacks be identified and implemented by students with minimal security training?

The research contribution is an initial empirical analysis of the viability of compromising a medical training mannequin along with providing the foundation for future research in this area. The paper is structured as follows: Section two discusses relevant medical device research. Section three presents the methodology and the experimental design implemented in this study. Section four discusses the implementation and the results. Section five draws conclusions and section six presents future work.

## Relevant Work

Recent interest in pervasive healthcare technology instigated a multitude of research efforts that include, but are not limited to, the security of implantable devices (Halperin et al. 2008b; Ransford et al. 2014), wireless healthcare devices (Al–Busaidi and Khriji 2014; Malasri and Lan 2009), and wearable medical devices (Ya-Li et al. 2014) along with wireless body and sensor networks (Anagnostopoulos et al. 2014; Diallo et al. ; Manfredi 2014). Ransford, et. al., (2014) applies fundamental security concepts to the development of implantable medical device domains. The authors acknowledge that there is tension between security solutions and implanted medical devices. These tensions include a) the need to fail 'open' for medical emergency access versus being constantly secure, and b) the need for security encryption versus the impact on functionality and energy consumption.

Malasri, et. al., (2009) examines the security threats facing implantable devices and classifies implantable wireless devices into three categories that include "identification, monitoring, and control" devices to discuss security issues. Identification simply provides identifiable information about an individual and is subject to tracking, harvesting, relay, cloning and physical attacks. The authors define monitoring devices as devices that measure patient physiological information. The authors note that the attacks acknowledged are similar to that of identification with the addition of medical identity theft and denial of service. The control classification involves devices that are capable of impacting a patent's physiological characteristics. This includes devices that dispense drugs like an insulin pump or devices that regulate organs like heart and brain pacemakers. The authors note that the unique threat in this category is the ability to wirelessly control or reprogram the device, potentially causing physical damage. The authors indicate that there is a lack of research focusing on vulnerabilities in reprogrammable control devices.

Li, et. al.'s, (2014) research into attacking a glucose monitor indicates that data is being passed in the clear. They report that being able to successfully acquire the device PIN for the glucose meter and monitor, along with generating and transmitting accepted data packets to the insulin pump that contained incorrect information, potentially, leads to unacceptable consequences.

Halperin, et. al., (2008b) propose a security and privacy evaluation framework for future wireless implantable medical devices. The paper outlines basic security concerns along with acknowledging conflicts. Those conflicts include the implementation of encryption and the impact on longevity and performance, increased wireless communications potentially increases attack exposure. They acknowledge that their solution to security problems for implantable medical devices ultimately involves collaboration between industry, security and medical communities in terms of both policies and devices.

Halperin, et. al. (2008a) takes a more in-depth look at pacemakers. They were able to intercept, understand and extract information from RF transmissions between the device and the programming unit. The eavesdropping allowed them to gather data on the device, the name of the patient and the patient history. They also indicate that they were able to modify the behavior of the device.

Gollakota, et. al., (2011) explores a solution for non-invasive secure implantable medical devices that shields the device by jamming outgoing and incoming signals. The idea is to place the security responsibilities on a separate device that acts as a gateway for authorized devices. The authors acknowledge that a sufficiently high powered broadcast is a limitation to their solution. They did counter this issue by integrating an alarm when a high powered broadcast is detected.

Diallo, et. al., presents an architectural solution that combines a wireless body area network with modeling techniques to provide storage infrastructure and improve query processing times. It is interesting to note that the data being collected for the body like blood pressure, heartbeat, insulin levels and body temperature are transmitted to a mobile device that establishes a secure connection to the storage area and then transfers the data. Previous research by Glisson, et. al., (2013) (2011) demonstrates that mobile devices retain residual data when interacting with users or software.

Previous medical device research has identified security issues, deliberated challenges and proposed solutions. However, there is minimal empirical research investigating the identification of security issues in medical training environments.

# Methodology

This research investigates the viability of compromising a mannequin system used in medical simulation training environments. The idea focusses on the network communication between the medical mannequin and the controlling computer. A team of fourth year undergraduate computing students were assembled and given access to a medical training mannequin that is currently deployed as an active medical training device in the College of Nursing at the University of South Alabama. The students were given full freedom to evaluate and exploit any identified vulnerabilities in either the software or communication mechanisms of the system itself. It should be noted that the experiment took place on a production device during an academic semester and that no modifications were made to the medical mannequin prior to allowing students to perform an analysis.

## *Experiment Configuration*

The student team had freedom to choose any network traffic capture tool for their study. The tools and environment used by the students included a Lenovo attack laptop running Microsoft Windows 8.1 Pro, Sun Virtual box (version: 4.3.8) with BackTrack 5 Release 3, iStan medical mannequin, iStan laptop running OSX Lepord (version: 10.5.2), iStan Muse software (version: 2.1), and a monitor used to display the mannequin's vitals to the medical trainees utilizing Touch Pro display software 2.0

The iStan is a medical mannequin that simulates respiratory, neurological and cardiovascular systems (CAE Healthcare). In other words, the iStan simulates a living or, as the case may be, a dying person in that it speaks, bleeds from two locations, secretes bodily fluids, has a blood pressure, a heart rate, and breathes along with having realistic airways and joints. The iStan is shown in Figure 1 - iStan. The interface for the Muse software that interacts with the iStan is shown in Figure 2 – Muse software.

During typical training sessions and examinations used by the College of Nursing, the Muse software allows an operator to control the iStan mannequin remotely by applying pre-loaded scenarios and ad-hoc inputs that represent real life medical situations. Nursing students make observations of vital signs and physical responses of the iStan mannequin as they would a normal patient. Scenarios dictate a certain prescribed set of responses by the nursing students, which provides a means of feedback and evaluation.

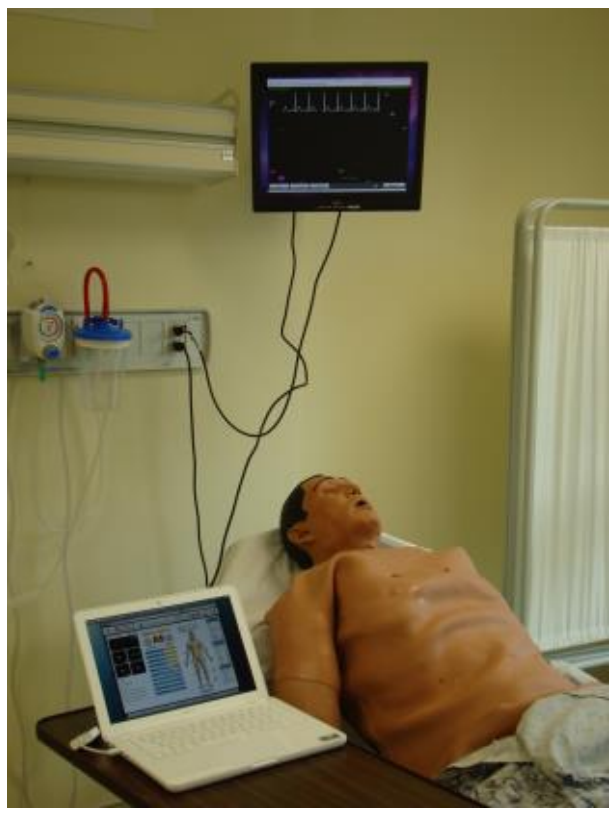

Figure 1. iStan

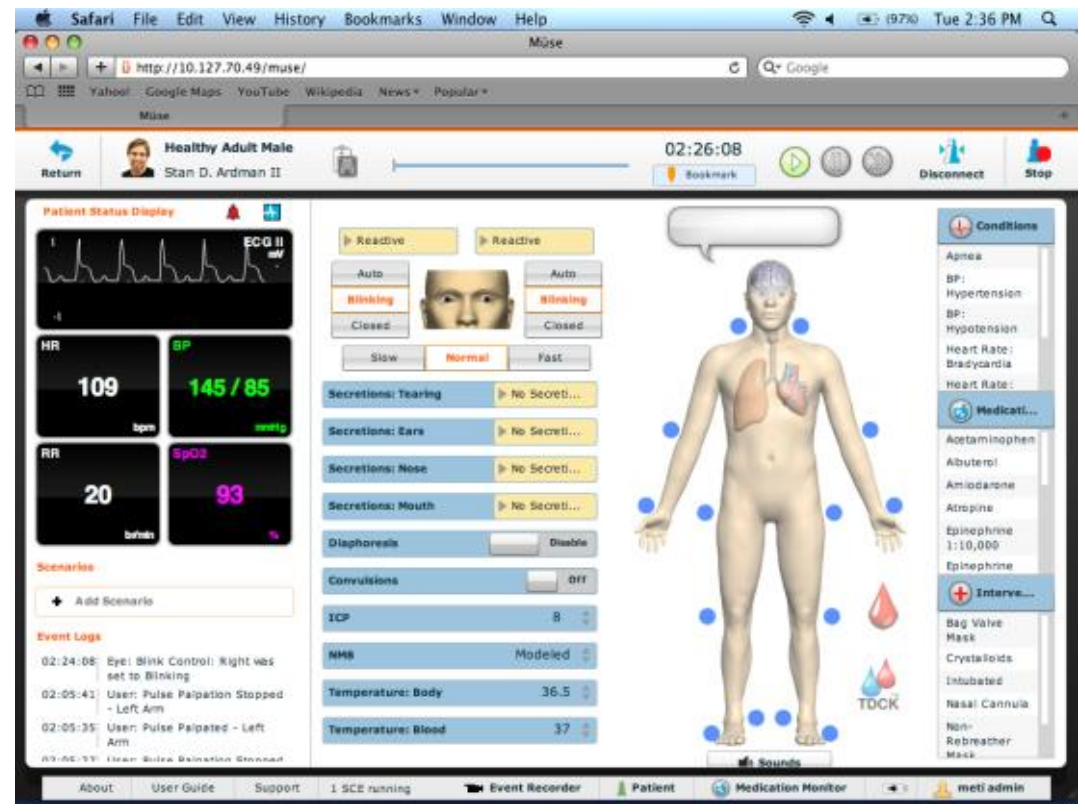

Figure 2. Muse software

### *Assessment and Exploitation*

As a **first step** in the analysis, students consulted publically available iStan device documentation to identify direct and indirect dependencies. Students looked for architectural and configuration information that might identify potential areas of weakness that could be easily attacked based on the publicly available documentation. According to the user manual available from CAE Healthcare (CAE Healthcare), the iStan Medical Training System software can be implemented on Macintosh or Microsoft Windows operating systems.

The iStan's front-end platform utilizes Adobe Flash Player and Muse software, which is a browser based application, to interact with iStan's profiles. The network protocols used by the iStan consist of TCP and 802.11 wireless transmissions between the controlling laptops and the iStan's internal access point. The iStan's direct dependencies consist of either a Windows or OS X machine, iStan mannequin and a properly configured access point. iStan's indirect dependencies are properly trained nursing students and nursing professors.

For the **second step**, students expanded on the identified weaknesses in the first step by selecting known attacks that take advantage of identified weaknesses. These included brute force attacks, a Denial of Service (DoS) attack, and security control attacks. The first simple attack identified was to compromise a training scenario by instigating a DoS attack using HPING3, a testing tool available in Linux and Windows environments (The Sprawl). As a second attack, students chose the open source software known as Reaver, which is included with BackTrack 5. Reaver allows a brute force attack against Wi-Fi Protected Setup register PIN numbers (Reaver WPS ; Viehböck 2011).

In the **third step**, students investigated the configuration of the attack approach by looking at published iStan documentation and available attack tools. They examined the implementation using either a live CD or a virtual environment. Due to practical considerations and time constraints, students focused on selecting and implementing tools that are relatively easy to acquire and utilize. These first three steps set the stage for actual attacks.

In the **fourth step** of methodology, students executed the individual attacks against the iStan mannequin. The initial attack in this stage consisted of a brute force attack against the device using Reaver and BackTrack 5. The results of the attacks were monitored, captured and inspected to ascertain the impact on the iStan device. The controlling medical mannequin's computer was also monitored to see if it provided any indications that there was something wrong.

BackTrack software was the only tool used for penetration testing even though additional customized attacks may have revealed more information. The attacks took place within a limited space consistent with an attacker who is in proximity to the medical training being conducted within the College of Nursing. Attack range variations were not examined. All other artifacts and interactions with the medical training mannequin were outside the scope of this project.

### Research Context

The focus of this research project was a proof of concept to show how focused attacks could be executed on a medical training mannequin. The scope of the experiment was limited to production deployed medical training mannequins that are currently in use at the University of South Alabama's College of Nursing.

A major goal of the research project was to assess whether or not students with basic information technology and computer science background could successfully accomplish medical device vulnerability assessment in a short amount of time. In our case, students had a semester (four months) to learn any necessary tools, security concepts, and penetration testing skills. They also had to research specific medical devices (like the iStan mannequin) and create realistic versions of potential attacks based on their research. Some students had prior coursework in network and operating system security, but no students had ever been taught penetration testing skills, specifically. Information assurance faculty members at the University served as mentors and sponsors of the project, but only provided the students broad parameters and access to necessary equipment in which to conduct the experiment. The project mentors provided little direct guidance on specific tools or techniques for the students to use other than providing resource links and reading materials.

The results of the project indicate that students with a computing curriculum background could learn security concepts and apply them in a focused setting, even if they are not specifically trained in penetration testing. The project also provided a template for future student research projects related to medical device security testing and established a general methodology which can be repeated.

## Implementation and Results

Given the time constraints of the project and the convenient availability of open source tools such as BackTrack, students were able to conduct brute force attacks and denial of service attacks against two different iStan mannequins.

### Live CD brute force attack

The implementation procedures for the brute force attack using BackTrack 5 from a live CD are provided below.

1. Burn ISO image of BackTrack 5 release 3 to DVD. (http://www.backtrack-linux.org/backtrack/backtrack-5-r3-released/)
2. Insert ISO DVD into suitable computer and restart the computer with disk in tray.
3. Enter the boot menu screen (black screen with boot medium options) and select boot from disk tray.
4. After BackTrack 5 loads and command screen is fully initiated, type in the command 'start x'. This will load the Graphic User Interface for BackTrack 5.
5. Under the applications tab at the top of the desktop select command prompt.
6. Under command prompt type in 'iwconfig', this shows the wireless card that is available on the machine. Results are illustrated in Figure 3 - Iwconfig.
7. At the command prompt 'airmon-ng start wlan0' was entered. This places the wireless card into monitor mode so it can scan for available wireless access points. Results of this command are seen in Figure 4 – Wireless card.
8. Typing 'airodump-ng mon0' at the command prompt shows all of the available access points within range of the computer as displayed in Figure 5 – Access point.

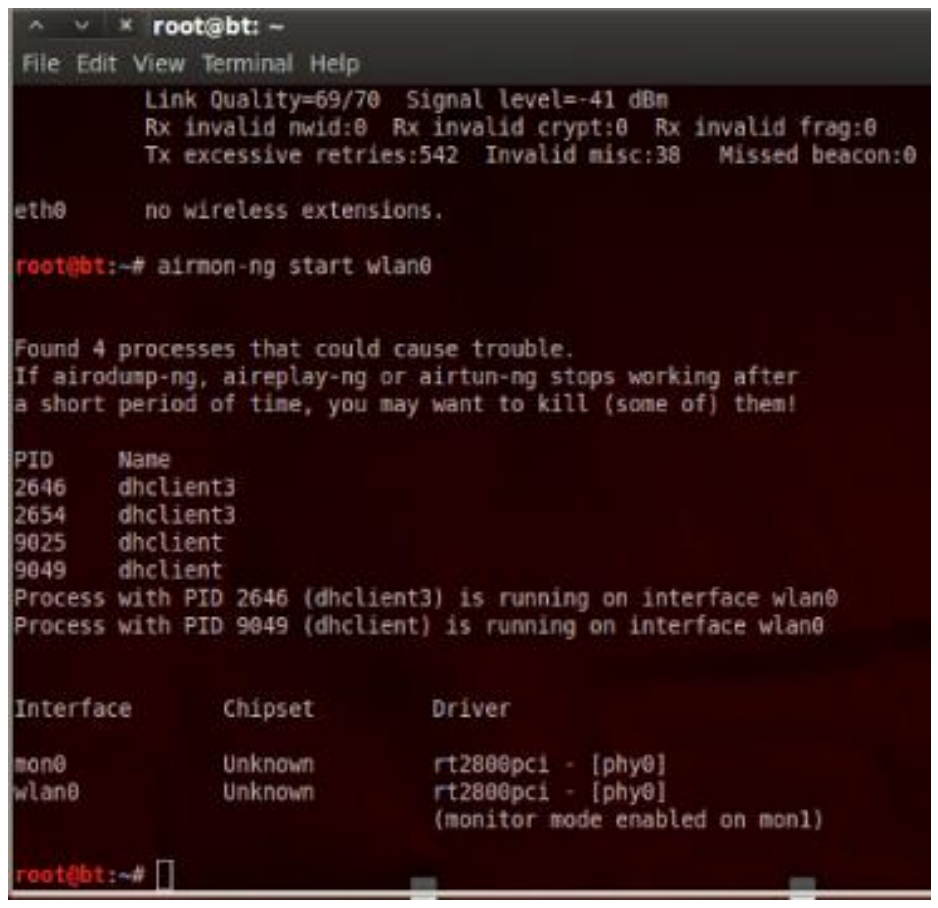

Figure 3. Iwconfig

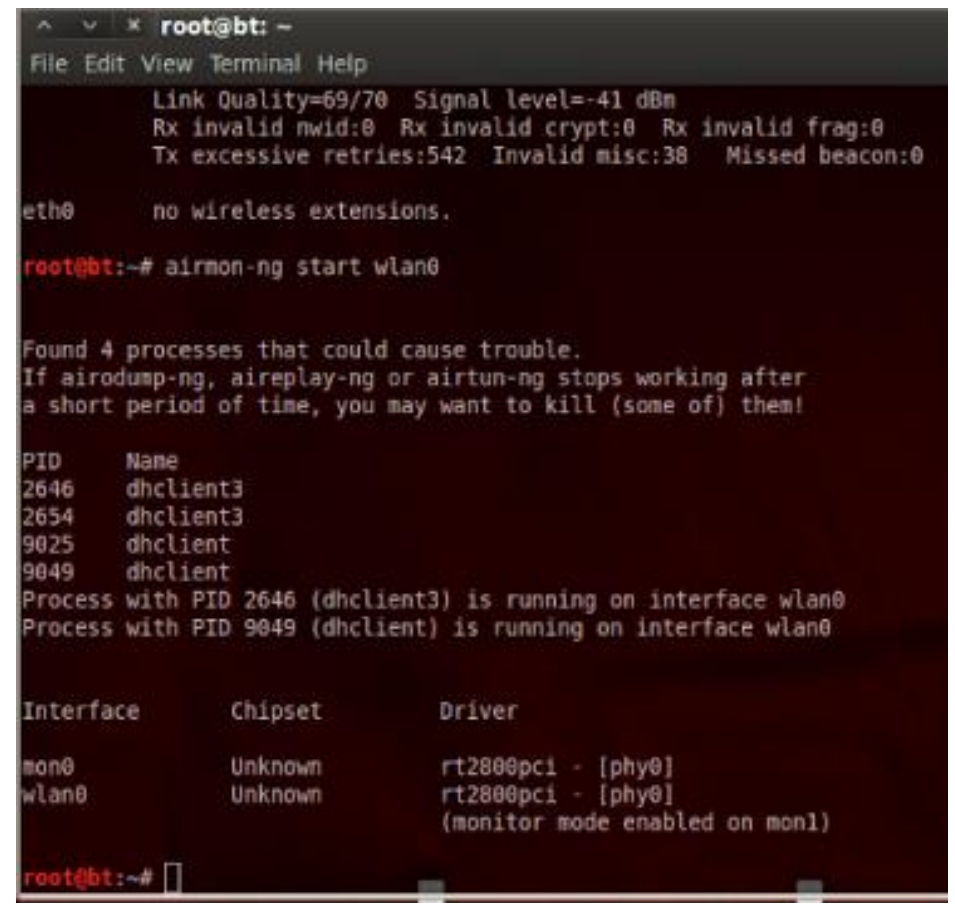

Figure 4. Wireless card

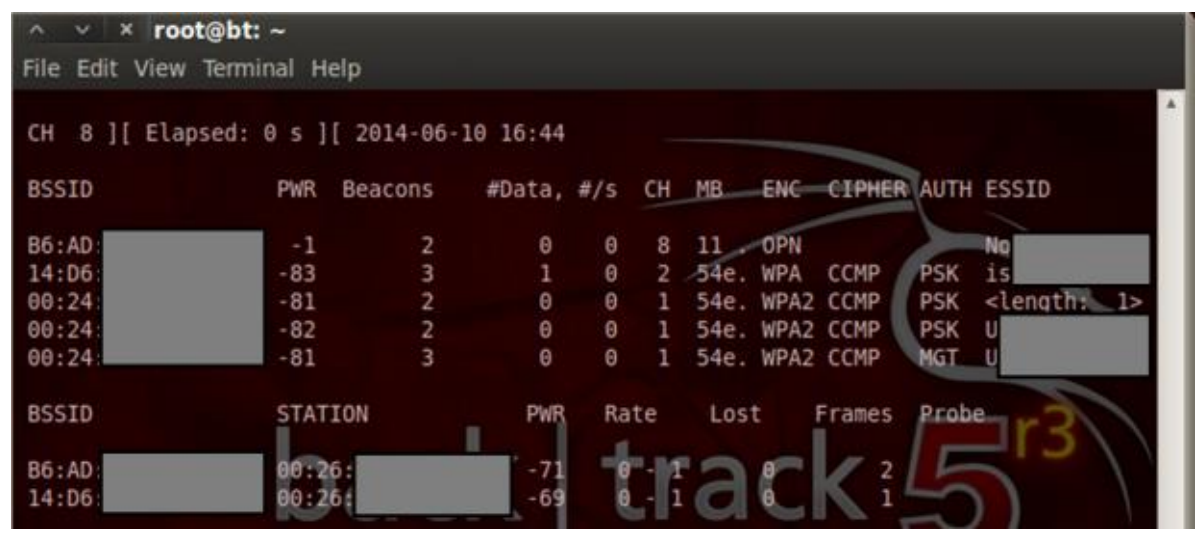

Figure 5. Access points

9. Identify the iStan machine that will be utilized for the attack. In this case, it is iSXXXXX. The MAC address for this device is 14:D6:XX:XX: XX:XX and channel number that the access point is running on is CH 2.
10. Type in command prompt reaver -i <moninterface> -b <MacAddress> -c <channel number> -d 0
    a. <moninterface> is the name of the wireless card: mon0
    b. <MacAddress> is the mac address for the Istan's access point: 14:D6:XX:XX:XX:XX
    c. <channel number> channel number of istan's AP of which its running on: 2
11. Command line used for this attack is as follows:
12. reaver -i mon0 -b 14:D6:XX:XX:XX:XX -c 2 -d 0
13. The results are displayed in Figure 6 - Reaver results live CD. It is interesting to note that it took a little over 7 hours to complete. The last 3 lines of the command below state as follows:
    a. WPS PIN: '33XXXXXX' is the PIN # for the access point
    b. WPA PSK: 'isXXXXXX' is the password for the access point.
    c. AP SSID: 'isXXXXXX' is the name of the Access point.

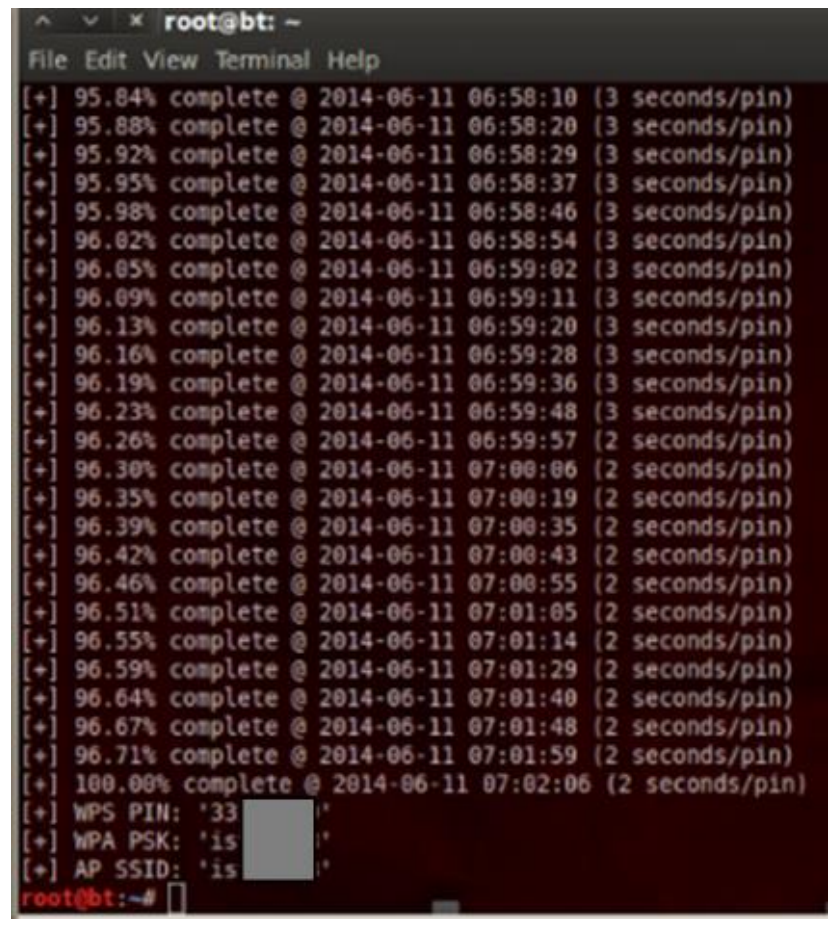

Figure 6. Reaver results live CD

## *Virtual machine brute force attack*

For comparison purposes, the same attack was repeated on another iStan using a virtual machine configuration. Several configuration steps were introduced at the beginning of the experiment which included:

1. An instance of Virtual Box 4.3.8 was installed on a Lenovo T430 machine running Windows 8.1.
2. Virtual Box extension package for USB compatibility was installed on this instance.
3. BackTrack 5 Release 3 was installed into virtual environment.
4. The USB interface with a virtual machine was configured to accept an external wireless adapter.
5. The installed and configured instance of the virtual machine was then started.

Once these steps had been introduced, the steps from the previous experiment were repeated from step six. The iStan used for this experiment was identified as iSXXXXX. The command to initiate the Reaver program was entered as follows: reaver -i mon0 -b 14:D6:XX:XX:XX:XX -c 3 -d 0. This is demonstrated in Figure 7.

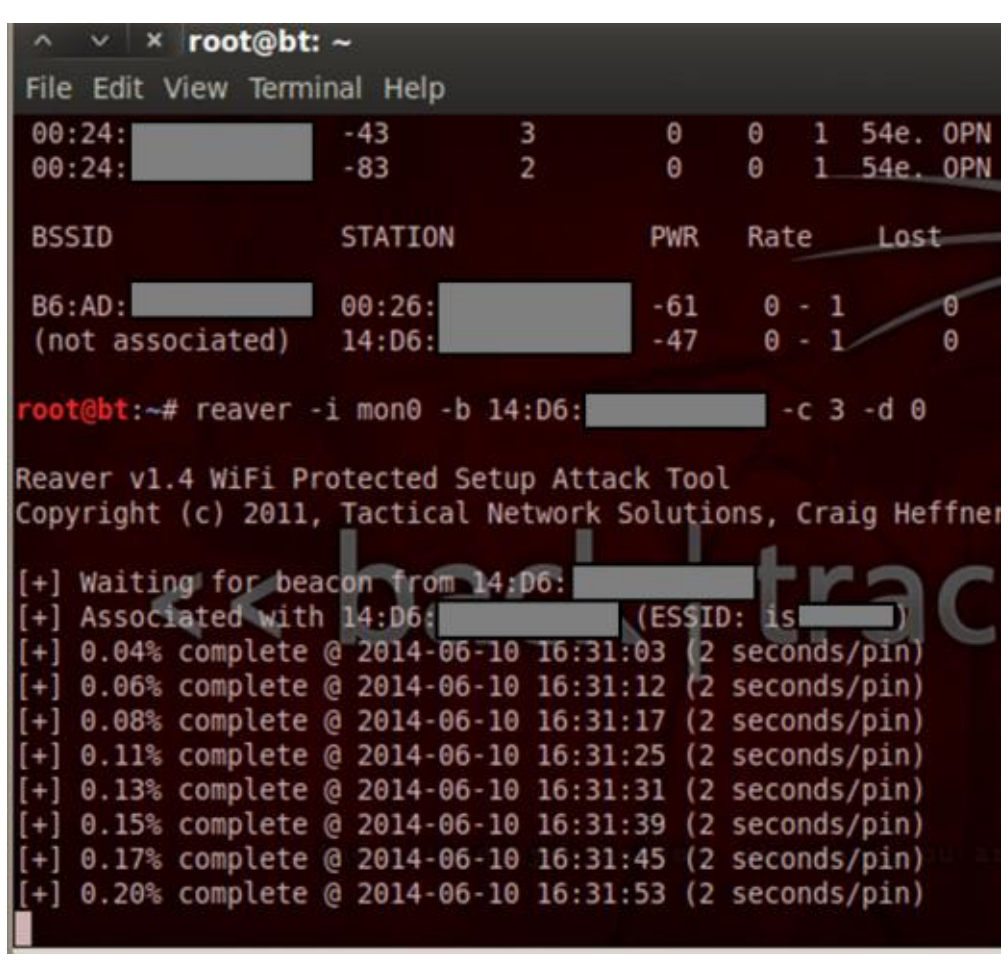

Figure 7. Reaver execution virtual machine

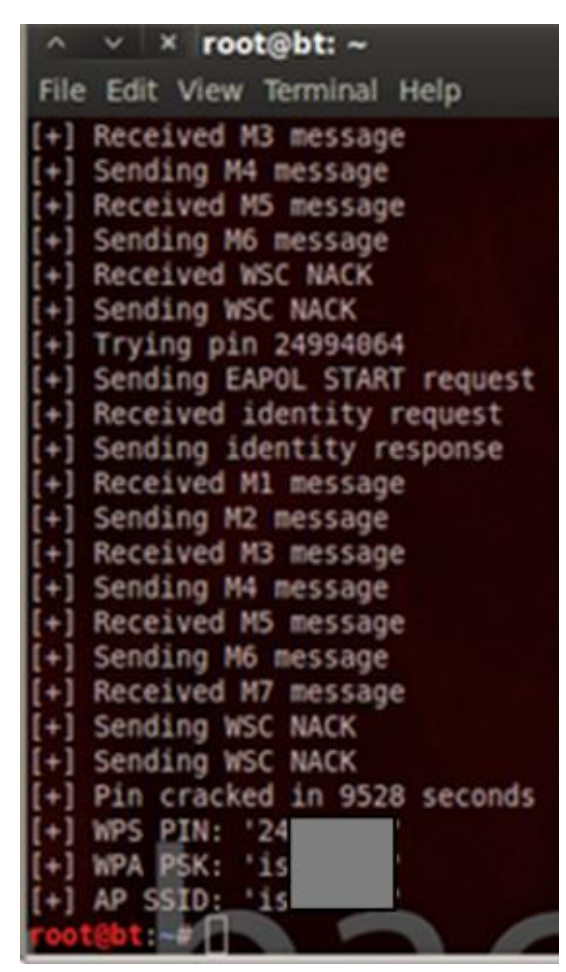

Figure 8. Reaver results virtual machine

It took 9,528 seconds or two hours thirty-eight minutes and forty-eight seconds to ascertain the passphrase for this device. This information recovered is displayed in Figure 8 and summarized below:

- WPS PIN: '24XXXXXX' is the PIN number for the access point

- WPA PSK: 'isXXXXXX' is the password for the access point (same as id)
- AP SSID: 'isXXXXXX' is the name of the Access point.

## *Denial of service attack*

For the denial of service attack, the first seven procedural steps and the results from implementing these steps using BackTrack 5 are the same as the brute force attack. Step 8 is where the process starts to deviate from the previous attack.

8. Type 'airodump-ng mon0' into command prompt. This will show all of the available access points within range of the computer. The access points that were available are displayed in Figure 9.
9. The access point isXXXXX was chosen for this attack: (14:D6:XX:XX:XX:XX channel 3). The following is a screenshot of Muse software that is used to communicate and control the iStan medical mannequin.

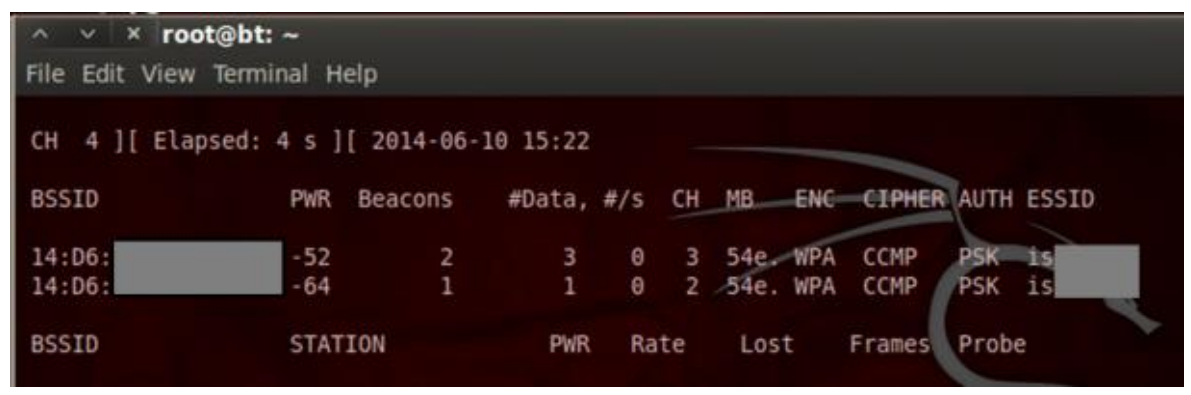

Figure 9. Denial of service attack access points

10. Type command 'echo 14:D6:XX:XX:XX:XX > blacklist' , This command saves the Mac address as a string to a file called blacklist
11. Type command 'mdk3 <monitor interface> d -b <file name that contains MAC address> -c <Channel number>':
    - Where <monitor interface> is "mon0" as listed in step 7
    - Where <file name that contains MAC address> is "blacklist"
    - Where <Channel number> is "3" as listed from step 8. For this attack the following command was entered: mdk3 mon0 d -b blacklist -c 3. In this attack, the command takes the saved Mac address from the blacklist file and uses a script called mdk3 to send disconnect signals to the access point on iStan. Figure 10 - Denial attack illustrates the results of running the command.

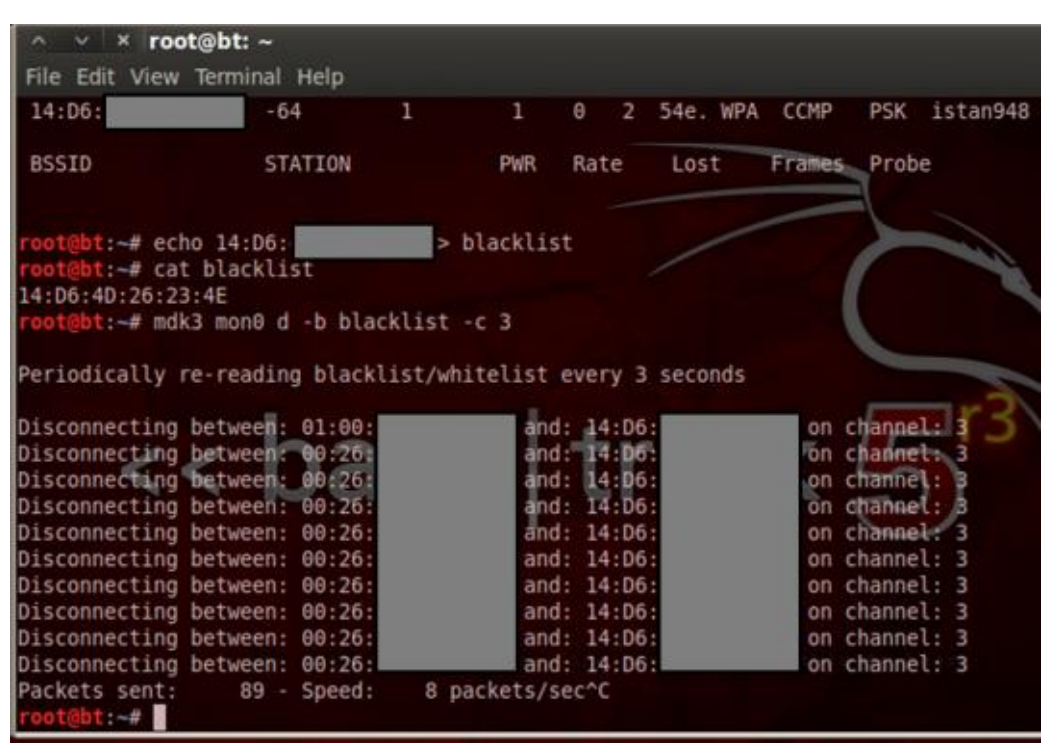

Figure 10. Denial attack

Figure 11 illustrates that the interface to the Muse software displayed a connection error message during the DOS attack. The program was not able to send any setting changes to the iStan medical mannequin.

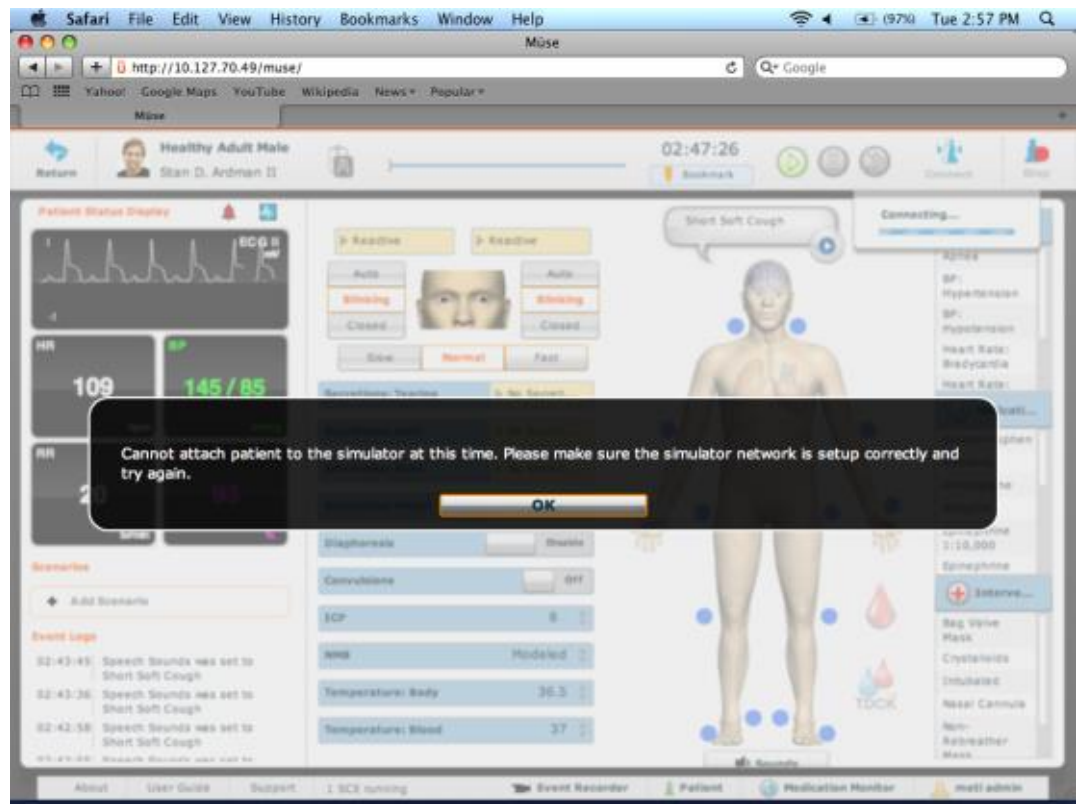

Figure 11. Connection error verification

## Conclusions

Medical equipment is becoming progressively dependent on technology inherently increasing the challenges and complexity involved with teaching, learning, and investigating this equipment. The impact is, potentially, compounded when breaches are considered in training environments. The results of this proof of concept research supports the hypothesis that medical training mannequins are at risk. The data demonstrates that it is possible to launch a successful attack against a medical mannequin in a production environment.

Based on traditional cryptographic properties, the differences in completion times for the two attacks is expected. The nature of brute force attacks leads to testing, on average, half of the key space creating vast discrepancies in completion times. However, future research should investigate the integration of customized dictionaries in order to reduce the overall time associated with the brute force attack. Terms for the custom dictionaries can be derived from the user guide associated with different mannequin and training facility information. This may also be mitigated in the future with more sophisticated tools or tailored algorithmic solutions. The tools implemented in this experiment were not sophisticated or difficult to acquire. This alone highlights the lack of security that is currently being implemented in these devices.

The components of the medical mannequin identified as vulnerable to attack in this research include the network security solution and the network protocol. The security solution was breached using an open source brute force attack against the router Personal Identification Number (PIN). The network protocol was attacked through a denial of service attack. For security reasons, the actual pins and full device MAC addresses are not displayed in the paper.

Collaboration between the School of Computing and the College of Nursing at the University of South Alabama will seek to integrate cyber-based scenarios into nursing education scenarios. As in the other security contexts, security can be viewed as a subset of reliability. In many cases failure could be due to device or malfunction or from byzantine actions of a malicious adversary. By integrating cyber-based scenarios into medical training, the College of Nursing foresees that future practitioners will be trained to deal with medical device failures, byzantine or otherwise, and will reinforce the use of alternate or traditional techniques that do not rely on technology.

If medical training environments are breached, the long term ripple effect on the medical profession, potentially, impacts thousands of lives due to incorrect analysis of life threating critical data by medical personnel. This research provides the foundation for investigating the development and establishment of a methodology for performing vulnerability assessment and penetration testing. It also provides insight into the integration of these results into real-world educational programs and curriculums benefiting computer professionals, medical practitioners and academicians.

## Future Work

In addition to the attacks discussed in this paper, future research will consider the implementation of more complex security issues like the development and propagation of software designed to interfere with the operation of medical devices. Once this has been achieved, the detection of this type of breach will need to be researched.

In conjunction with software research, medical device studies need to investigate the applicability of policies, standards and procedures. Research should also investigate reverse engineering mobile medical devices to ascertain security vulnerabilities, establish residual data that is resident on the device after various interactions along with establishing proper forensic procedures for investigating mobile medical embedded and wearable devices. This area of research also presents a unique opportunity to interface with and impact medical training environments that are dependent on technology.

Hence, future work will identify residual data that results from iStan breaches that would be beneficial in a digital forensic investigation. Other manufacturer's makes and models of medical training mannequins along with implantable devices, such as pace makers and defibrillators, need to be researched from a digital forensics perspective. These research ideas also warrant an examination of the equipment used to train doctors to successfully implant these devices. The idea is to progress a general methodology that can be implemented in future investigations of medical equipment. This methodology can also be used as a guide for security equipment validations to minimize the risk of using compromised equipment in a production environment.

In addition to being sure that the equipment has not been compromised, research needs to examine the dependency of medical professionals on technology output. This will involve running studies to see how medical professionals cope when technology has been removed from their environment. This research should also examine technology dependence from the perspective of different countries and associated cultures. Ultimately, the goal is to develop relevant curriculum input that implements imperfect technology feedback to train medical personnel to recognize issues along with interpreting and interacting with flawed data effectively and efficiently.

## References

Al–Busaidi, A. M., and Khriji, L. 2014. "Wearable Wireless Medical Sensors toward Standards, Safety and Intelligence: A Review," *International Journal of Biomedical Engineering and Technology* (14:2), pp. 119-147.

Anagnostopoulos, C., Hadjiefthymiades, S., Katsikis, A., and Maglogiannis, I. 2014. "Autoregressive Energy-Efficient Context Forwarding in Wireless Sensor Networks for Pervasive Healthcare Systems," *Personal and Ubiquitous Computing* (18:1), pp. 101-114.

Basu, E. 2013. "Hacking Insulin Pumps and Other Medical Devices from Black Hat." Retrieved May 1, 2014, from http://www.forbes.com/sites/ericbasu/2013/08/03/hacking-insulin-pumps-and-other-medical-devices-reality-not-fiction/

CAE Healthcare. "Istan." Retrieved June 4, 2014, from http://www.caehealthcare.com/eng/patient-simulators/istan

Diallo, O., Rodrigues, J. J. P. C., Sene, M., and Niu, J. "Real-Time Query Processing Optimization for Cloud-Based Wireless Body Area Networks," *Information Sciences*:0).

Federal Bureau of Investigation. 2014. "Fbi Cyber Division Bulletin: Health Care Systems and Medical Devices at Risk for Increased Cyber Intrusions." Retrieved June 4, 2014, from https://publicintelligence.net/fbi-health-care-cyber-intrusions/

Filkins, B. 2014. "Health Care Cyberthreat Report," SANS, p. 41.

Fu, K., and Blum, J. 2013. "Controlling for Cybersecurity Risks of Medical Device Software," *Commun. ACM* (56:10), pp. 35-37.

Glisson, W. B., and Storer, T. 2013. "Investigating Information Security Risks of Mobile Device Use within Organizations " in: *Americas Conference on Information Systems (AMCIS)*.

Glisson, W. B., Storer, T., Mayall, G., Moug, I., and Grispos, G. 2011. "Electronic Retention: What Does Your Mobile Phone Reveal About You?," *International Journal of Information Security* (10:6), pp. 337-349.

Gollakota, S., Hassanieh, H., Ransford, B., Katabi, D., and Fu, K. 2011. "They Can Hear Your Heartbeats: Non-Invasive Security for Implantable Medical Devices," *SIGCOMM Comput. Commun. Rev.* (41:4), pp. 2-13.

Halperin, D., Heydt-Benjamin, T. S., Ransford, B., Clark, S. S., Defend, B., Morgan, W., Fu, K., Kohno, T., and Maisel, W. H. 2008a. "Pacemakers and Implantable Cardiac Defibrillators: Software Radio Attacks and Zero-Power Defenses," in: *Proceedings of the 2008 IEEE Symposium on Security and Privacy*. IEEE Computer Society, pp. 129-142.

Halperin, D., Kohno, T., Heydt-Benjamin, T. S., Fu, K., and Maisel, W. H. 2008b. "Security and Privacy for Implantable Medical Devices," *Pervasive Computing, IEEE* (7:1), pp. 30-39.

Industrial Control Systems Cyber Emergency Response Team. 2013. "Alert (Ics-Alert-13-164-01) Medical Devices Hard-Coded Passwords." from http://ics-cert.us-cert.gov/alerts/ICS-ALERT-13-164-01

Lee, T. 2014. "Hackers Break into Networks of 3 Big Medical Device Makers." Retrieved May 1, 2014, from http://www.sfgate.com/news/article/Hackers-break-into-networks-of-3-big-medical-5217780.php

Lemos, R. 2013. "Medical-Device Flaws Will Take Time to Heal." Retrieved May 1, 2014, from http://www.darkreading.com/medical-device-flaws-will-take-time-to-heal/d/d-id/1140255?

Li, C., Zhang, M., Raghunathan, A., and Jha, N. 2014. "Attacking and Defending a Diabetes Therapy System," in *Security and Privacy for Implantable Medical Devices*, W. Burleson and S. Carrara (eds.). Springer New York, pp. 175-193.

Malasri, K., and Lan, W. 2009. "Securing Wireless Implantable Devices for Healthcare: Ideas and Challenges," *Communications Magazine, IEEE* (47:7), pp. 74-80.

Manfredi, S. 2014. "Congestion Control for Differentiated Healthcare Service Delivery in Emerging Heterogeneous Wireless Body Area Networks," *Wireless Communications, IEEE* (21:2), pp. 81-90.

Neely, M. 2013. "Top Five Vulnerabilities Hackers Can Use Right Now to Shut Down Medical Devices." Retrieved May 1, 2014, from http://blog.securestate.com/healthcare-interrupted/

Pauli, D. 2013. "Patient Data Revealed in Medical Device Hack." Retrieved May 1, 2014, from http://www.itnews.com.au/News/329222,patient-data-revealed-in-medical-device-hack.aspx

Ransford, B., Clark, S., Kune, D., Fu, K., and Burleson, W. 2014. "Design Challenges for Secure Implantable Medical Devices," in *Security and Privacy for Implantable Medical Devices*, W. Burleson and S. Carrara (eds.). Springer New York, pp. 157-173.

Reaver WPS. "Reaver Wps Brute Force Attack against Wifi Protected Setup." Retrieved June 11, 2014, from http://sourceforge.net/projects/reaver-wps.mirror/

Storm, D. 2013. "Fda Asks Hackers to Expose Holes in Medical Devices, but Many Researchers Fear Cfaa & Jail." Retrieved May 1, 2014, from http://blogs.computerworld.com/cybercrime-and-hacking/22529/fda-asks-hackers-expose-flaws-medical-devices-researchers-fear-cfaa-prison

The Sprawl. "Researchhping." Retrieved June 9, 2014, from https://thesprawl.org/research/hping/

U.S. Food and Drug Administration (FDA). 2013a. "Content of Premarket Submissions for Management of Cybersecurity in Medical Devices - Draft Guidance for Industry and Food and Drug Administration Staff." Retrieved May 1, 2014, from http://www.fda.gov/

U.S. Food and Drug Administration (FDA). 2013b. "Fda Safety Communication: Cybersecurity for Medical Devices and Hospital Networks." Retrieved May 1, 2014, from http://www.fda.gov/medicaldevices/safety

United States Government Accountability Office Report. 2012. "Medical Devices Fda Should Expand Its Consideration of Information Security for Certain Types of Devices."

Viehböck, S. 2011. "Brute Forcing Wi-Fi Protected Setup: When Poor Design Meets Poor Implementation.", from http://sviehb.files.wordpress.com/2011/12/viehboeck_wps.pdf

Weaver, C. 2013. "Patients Put at Risk by Computer Viruses." Retrieved May 1, 2014, from http://online.wsj.com/article/SB10001424127887324188604578543162744943762.html

Wilson, T. 2013. "Medical Devices Subject to Cyberattack, Fda Warns." Retrieved May 1, 2014, from http://www.darkreading.com/medical-devices-subject-to-cyberattack-fda-warns/d/d-id/1139959?

Ya-Li, Z., Xiao-Rong, D., Poon, C. C. Y., Lo, B. P. L., Heye, Z., Xiao-Lin, Z., Guang-Zhong, Y., Ni, Z., and Yuan-Ting, Z. 2014. "Unobtrusive Sensing and Wearable Devices for Health Informatics," *Biomedical Engineering, IEEE Transactions on* (61:5), pp. 1538-1554.